# Design and validation of an instrument to test students' understanding of the apparent motion of the Sun and stars

Hans Bekaert,[1] Hans Van Winckel,[2] Wim Van Dooren,[3]
An Steegen,[4] and Mieke De Cock[1,*]

[1]*KU Leuven, Department of Physics and Astronomy and LESEC,
Celestijnenlaan 200c—box 2406, 3001 Leuven, Belgium*
[2]*KU Leuven, Department of Physics and Astronomy,
Celestijnenlaan 200d—box 2401, 3001 Leuven, Belgium*
[3]*KU Leuven, Centre for Instructional Psychology and Technology,
Dekenstraat 2—box 3773, 3000 Leuven, Belgium*
[4]*KU Leuven, Department of Earth and Environmental Sciences & LESEC,
Celestijnenlaan 200e—box 2409, 3001 Leuven, Belgium*

Young children, students, and adults may have alternative ideas about the motion of the Sun and stars as we observe them in the sky. However, a good understanding of this apparent motion is essential as a starting point to study more advanced astronomical concepts, especially when these include astronomical observations. In this paper, we describe the development and validation of the apparent motion of Sun and stars (AMoSS) test, which can measure to what extent students have insight into the apparent motion of the Sun and stars. We propose a framework that allows one to compare students' understanding of the specific aspects of these apparent motions in relation to the time of the day, time of the year, and the observer's latitude. For each of these aspects, we designed test items for both the Sun and the symmetric apparent motion aspect of the stars. The reliability and validity of the test are established by analyzing answers of both secondary school and university students and by presenting the questions to a panel of experts. We report on the design and validation process and present the final version of the test.

DOI: 10.1103/PhysRevPhysEducRes.16.020135

## I. INTRODUCTION AND PROBLEM STATEMENT

In the context of our project on the study of learning opportunities in a planetarium environment (SLOPE), we aim at developing a research-based planetarium presentation for 5th year secondary school students (16–17 year olds). In this article we report on the first step of this development process, which is the design and validation of a test to measure students' understanding of specific elements of the apparent motion of the Sun and stars.

Understanding the celestial motion of the Sun, Moon, and stars from the point of view of an observer on Earth, can be regarded as the starting point to develop insight into very basic astronomical phenomena such as the day-night cycle, the lunar phases, and the rising and setting of the Sun. A lack of understanding of the concepts underlying these basic phenomena prevents students from understanding more advanced astronomical topics [1].

However, research has shown that these basic concepts are difficult to grasp not only for young children but also for secondary school students, university students, and adults [2–7]. This can be attributed to the fact that children at a very young age (younger than 4 years) [8] already develop initial theories about the physical world in which they live. For example, they think that it will be night because the Sun sinks behind the mountains and then the Moon and stars rise [9]. Research into students' learning of science has shown that transforming these naive, initial ideas, which are based on everyday observations and experiences, into real scientific explanations is a slow and often unsuccessful process [10]. During this process, it is almost inevitable that misconceptions are formed: these are hybrid, intermediate states of understanding in which elements of the naive initial ideas are combined with elements of the correct scientific model, without, however, completely giving up those initial ideas while they are incorrect [10].

It is by now well established that only by means of very specific instruction, these alternative ideas can be turned into correct scientific knowledge [11,12].

In the context of our project, we focus on the apparent motion of the Sun and stars and the role a planetarium can play in the learning of these phenomena. There are at least two reasons why we have chosen to focus on these apparent motions.

First, although everybody has experience with the apparent motion of the Sun, several studies have shown that many people cannot really explain this phenomenon [13–16]. Moreover, although the apparent motion of the stars is in principle easier to comprehend, literature indicates that the opposite might be the case [14,15,17,18]. To our knowledge, student understanding of these apparent motions has not been studied and compared systematically. By limiting ourselves to the apparent motion of Sun and stars, we strive to obtain a deep, systematic insight into students' understanding of the differences and similarities between these celestial motions. At the same time, we aim at identifying the alternative models students use to explain their answers.

Second, our choice relates to the context of our research in which we study learning opportunities of planetarium visualizations. In their review of research on teaching astronomy in the planetarium, Slater and Tatge conclude that during the last decade the complex visualization possibilities of digital projection planetariums are improving learning gains [19]. However, as far as we know, no systematic study on the possible role of a planetarium in the study of the apparent motion of the Sun and stars has been carried out yet. Except for the research by Plummer [13,17] with primary and middle school children, the role a planetarium can play to scaffold deep understanding of these very basic ideas is not well understood.

In order to measure to what extent a series of lessons at school or a visit to a planetarium contributes to the development of a deeper insight into the apparent celestial motions, a valid and reliable test instrument is needed. Since such a test instrument is not available yet, we have developed one ourselves.

In this article, we describe the development of the test instrument about the apparent celestial motion of the Sun and stars, as well as its reliability and validity. Although we are interested in examining mental models students use while explaining their answers, we will reserve this for future work.

In Sec. II, we give a more detailed overview of students' difficulties and the associated proliferation of alternative ideas. We illustrate this with two case studies (seasons and the apparent motion of the Sun and stars). We elaborate on the link between a deep understanding of the apparent motion of the Sun and stars and the need for spatial thinking skills. We review the existing tests that are relevant for our purpose. Section III describes different steps of the AMoSS test development process and discusses the validation of the test. The last section concludes with a discussion and ideas for further research.

## II. BACKGROUND

### A. Students' difficulties in learning basic astronomical concepts

Several reviews on astronomy education [5,6] conclude that many phenomena in astronomy are difficult to grasp not only for students, but also for their teachers. Whereas the earlier studies mainly consisted of empirical attempts to describe students' alternative ideas, more recently the research became more theorized. Influenced by Vosniadou's work on conceptual change theory [15,20] attempts are made to identify in a more systematic way the mental models that underly the answers students give and the alternative ideas they express. Related to our research there are two domains of students' alternative ideas we want to highlight.

First, the "reasons for the seasons". In the video "A private universe" Schnepps and Sadler [21] demonstrated that even the brightest students fail to grasp this seemingly simple and fundamental phenomenon. Many studies have reported the same alternative ideas in a wide variety of different types of students of different ages and different levels [21–26]. The most common alternative idea is that it is warmer in summer than in winter because the Earth's orbit is highly elliptical, causing the Earth to be closer to the Sun in summer than in winter. As students learn about the relevance of the tilt of the Earth's axis, they integrate this new idea with what they already believe by assuming that the tilt brings one hemisphere closer to the Sun than the other. Some students also think the tilt of the Earth's axis changes during the year and that this change causes the seasons. Many students do not know that the place where the Sun sets varies throughout the seasons. A large percentage of students at all levels believe that the Sun can be seen overhead at noon every day. In his review study about learning about seasons [21] Sneider suggests that understanding the apparent motion of the Sun, based on daily observations and linked to the spinning of the Earth on its axis is a necessary first step in a possible successful learning progression for the seasons concept.

Second, many students have different alternative ideas precisely concerning the apparent motion of the Sun and stars. Vosniadou and Brewer [15] found that primary school children often think that stars are fixed and unmoving in the sky during the night, while they believe the Sun is "moving" during the day. In accordance with this conclusion Slater *et al.* reported that most undergraduate students in their study (38%) preferred the "fixed" notion of stars over "moving" stars at night [13]. Plummer found that 65% of her sample of eight graders described a fixed-star sky [14], while during her interviews all of the eighth-grade students demonstrated that the Sun rises and sets on opposite sides of the sky. The fact that students often reason differently about the apparent motion of the Sun and the apparent motion of the stars, reveals that they do not link these apparent motions to the rotation of the Earth around its axis. Heywood *et al.* [27] reported in their study about preservice teachers reasoning about the Sun's apparent motion that without exception the class ($n = 26$) attributed the day-night cycle to the Earth spinning on its axis, but there were no preservice teachers who related this explicitly to the Sun's apparent motion during the day.

There is clearly something inherently difficult in understanding these apparent motions. Several studies [2,3,21,28] suggest therefore that only by studying both the observations from Earth (geocentric frame of reference) and the actual motions as observed from an allocentric frame of reference, students can achieve the correct scientific insights. Only by learning to think and alter between these two frames of reference students can understand the apparent motion of the Sun and stars and link these to the actual motion of Earth. Probably, specific instructional strategies are needed to support students in this learning process.

Also, Cole *et al.* argue that even when a student holds accurate knowledge about the causation of astronomical phenomena, spatial thinking skills are needed to create an accurate mental model of these complex phenomena [29]. These involve highly complex cognitive activities [30]: to understand astronomical phenomena, one needs the ability to imagine objects from different perspectives and to track the motion of objects in multidimensional space [31]. One of the difficulties students have to deal with in astronomy lessons is that astronomical concepts are presented using a multitude of different disciplinary specific resources, including different representations, tools and activities [32]. Many of these discipline specific representations are one or two dimensional (e.g., maps, diagrams, graphs, etc.) and require the ability to extrapolate three dimensionality from them. Multiple studies have shown that spatial thinking skills contribute to students' performance in science, technology, engineering, and mathematics disciplines (biology, geology, physics, etc.) [33–35], but also that multidimensional thinking is challenging for many students [32].

In their study on the development and validation of a learning progression for change of seasons, solar and lunar eclipses and Moon phases, Testa *et al.* [3] confirm that spatial reasoning is a key factor for building an explanatory framework for celestial motion [1,27,36,37]. They also suggest that causal reasoning based on physics mechanisms underlying the astronomical phenomena, may significantly impact students' understanding. They argue that teaching celestial motions by using a learning progression research approach, which integrates causal reasoning with spatial thinking about the phenomena related to these motions, may help students progress from qualitative to more quantitative explanatory models about these motions.

In the context of our project, we are interested to look how a planetarium can contribute to this. As visualizing the (night) sky is one of the main goals of a planetarium, planetariums might be a powerful setting to support and enhance student learning of these astronomical phenomena. Yu [28,38,39] already highlighted the potential benefits of (digital) full dome astronomy education: revealing three-dimensional spatial relationships, demonstrating accurate motions of astronomical bodies. And while the traditional planetarium essentially provides a geocentric frame of reference, digital planetariums also use the available technology to contrast this geocentric frame with other frames of reference [6].

### B. Existing tests

In the literature, several tests exist that focus on a single astronomical topic. These so-called "concept inventories" are questionnaires with multiple-choice questions that focus on a single theme and for which the alternatives are based on research into students' thinking errors [40]. Examples are The Lunar Phases Concept Inventory [41], The Light and Spectroscopy Concept Inventory [42], The Astronomy and Space Science Concept Inventory [43], The Star Properties Concept Inventory [44].

In addition to these single-theme tests, there are also more general tests, such as the Astronomy Diagnostic Test (ADT) [45], the Test Of Astronomy Standards (TOAST) [46], The Astronomical Misconceptions Survey (AMS) [22] and the Introductory Astronomy Questionnaire (IAQ) [47], which gauge the astronomical knowledge of students more broadly. These are often used in a pretest and post-test setting to monitor student progress when taking an introductory course in astronomy.

Of all these tests, none is specifically about the apparent motion of the Sun and stars, and items are not systematically designed to provide a complete overview of students' understanding of these phenomena. Still, these more general tests contain a number of questions related to this theme and served as inspiration for our test. Questions about how the position of the Sun or stars in the sky changes after a period of one hour or month, can be found in the TOAST test [13] and the ADT test [45]. Questions about the cause of the seasons can be found in the AMS survey [22].

The analysis of the results of these questionnaires indicates that students often have an unclear understanding of the changes in the observable sky during a year [13,48,49]. However, neither the existing general tests nor the more specific concept inventories systematically probe students' understanding of the apparent motion of Sun and stars. As we want to gain insight in students' ideas of the different elements that underlie these apparent motions, we developed the AMoSS test.

### III. DEVELOPMENT OF THE TEST: METHOD AND RESULTS

Based on the recommendations of Adams and Wieman [50] on how to develop and validate instruments to measure learning of expertlike thinking, we developed the AMoSS test in three phases: (i) defining the specifications the test has to meet and delineating the scientific content that will be questioned, (ii) designing the questions, (iii) testing and optimizing the questionnaire with the target group in order to check whether the questionnaire meets the proposed specifications, and is reliable and valid.

### A. Specifications

We formulate the specifications of our questionnaire in detail:

1. The AMoSS test aims to test students' understanding of the apparent motion of the Sun and the stars during a day, a year and for different locations of the observer on Earth.
2. The AMoSS test should include the different elements that influence the observation of the apparent motion of the Sun and stars on Earth.
3. The AMoSS test must meet the basic principles of a good test [51]: The questions must be formulated unambiguously, the test must be valid and reliable.
4. Administering the AMoSS test with pen and paper should take a rather limited time (e.g., half an hour), so that the test can be used during a lesson at school or during a planetarium visit.
5. The AMoSS test must be written in the language of the target audience.
6. The AMoSS test must comply with a number of basic principles: each question may only address one concept and the alternatives should correspond with the most common alternative conceptions, as described in literature [22].
7. The AMoSS test must be easily adaptable for other target groups.
8. The AMoSS test should provide insight into the students' reasoning in order to be able to estimate which alternative ideas students have.

### B. Development process

In the next phase, we identified the elements that play a role in the apparent motion of the Sun and stars and we decided which to include in the test. We listed the major characteristics of the apparent motion of the Sun and stars (e.g., culmination height, position of the Sunrise and sunset, etc.) in relation to the time of the day, time of the year and position of the observer on Earth. Especially those characteristics which reveal differences and similarities between the apparent motion of the Sun and the apparent motion of stars, as seen by the observer on Earth, were brought together in a table.

This resulted in Table I with 7 categories (rows). It gives a symmetrical overview of elements related to the apparent motion of the Sun in the left column and the corresponding elements related to the apparent motion of stars in the right column. For each of these categories, we designed a parallel question for the Sun and the stars. We made sure the question formulation was as isomorphic as possible between the Sun and parallel star question (see examples in Fig. 1). Moreover, we paid special attention to the readability of the questions by limiting the number of words and omitting specialized jargon words. All questions are multiple choice and each question tests only one concept. The questions are written in such a way that with little effort a variant of the test can be made with the same questions but with an open answer format. The final version of the questionnaire is presented as Supplemental Material [52].

As mentioned in Sec. II. A on students' difficulties in learning astronomy some studies already probed for student understanding of certain aspects of the apparent motions. The following questions were inspired by the existing tests mentioned above, but all questions were reformulated and the figures were redrawn so that the assumed symmetry between the Sun and the stars is strongly expressed in the

TABLE I. Initial framework of the AMoSS test: Similarities and differences between the apparent motion of the Sun and stars.

| (I) Apparent motion of the Sun | (II) Apparent motion of a star |
|---|---|
| (A) Daily sun position changes: Sun's path. (Question I.A) | (A) Nightly star position changes: star trail. (Question II.A) |
| (B) Sun culmination changes during a year. (Question I.B) | (B) Star culmination does not change during a year. (Question II.B) |
| (C) Sunrise and sunset positions change during a year. (Question I.C) | (C) Star-rise and star-set positions do not change during a year. (Question II.C) |
| (D) Sun culmination depends on observer position. (Question I.D) | (D) Star culmination depends on observer position. (Question II.D) |
| (E) Sunrise and sunset position depend on observer position. (Question I.E) | (E) Star-rise and star-set position depend on observer position. (Question II.E) |
| (F) Speed of the apparent motion of the Sun changes during a year. (Question I.F) | (F) Speed of the apparent motion of a star does not change during a year. (Question II.F) |
| (III) Seasons: colder and warmer periods on a specific location during a year, due to Earth's revolution. (Question III) | (IV) Sky map changes on a specific location during a year, due to Earth's revolution. (Question IV) |

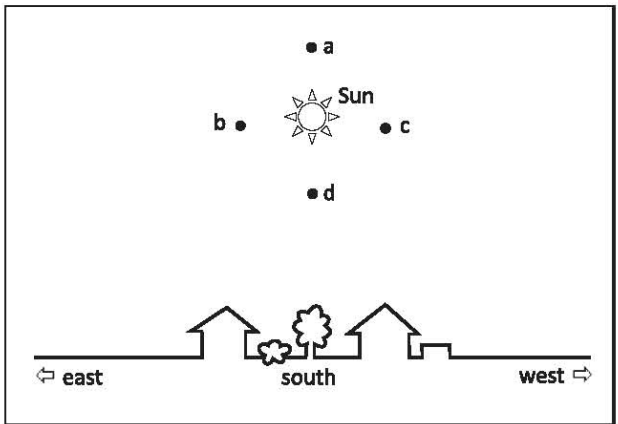 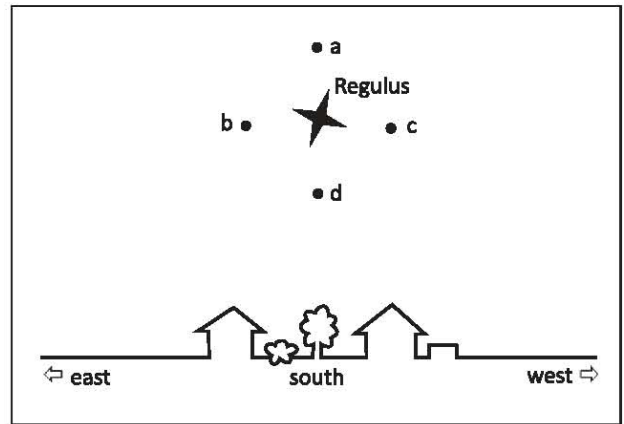

On March 21st, an observer in Brussels sees the Sun in the south high above the horizon as shown in the figure. Where does this observer see the Sun one hour later?

On March 21st, an observer in Brussels sees the star Regulus in the south high above the horizon as shown in the figure. Where will this observer see Regulus one hour later?

FIG. 1. The Sun and star question of category A are formulated as isomorphic as possible. See the Supplemental Material [52] for the complete version of the AMoSS test.

test (for an example see Fig. 2). For a complete comparison of the specific original test items with the redesigned AMoSS questions, we refer to the Supplemental Material [52] for more details:
- Question I.B is similar to question 4 of the TOAST test;
- Question I.C is similar to question 9 of the ADT test;
- Question III is similar to question 6 of the AMS survey;
- Questions II.A, II.B, and I.D are similar to questions that were also used in the ALIVE study [28].

With the aim of validation in mind, in order to get a good idea of the participants' reasoning, also a written explanation is asked for. In this explanation, the participant indicates, on the basis of a few sentences and possibly an accompanying sketch, the reasoning underlying his or her choice. This allows to check not only whether the question is understood as intended, but also to test whether a correct answer is based on a correct reasoning. Alternatively, for incorrect answers, we want to be able to probe the underlying alternative ideas.

We submitted the questionnaire to an expert panel consisting of a professor of physics education, a professor of geography education, a professor of educational psychology and a professor of astronomy. They were asked to answer the following questions:
- Does the test item question the intended concept?
- Will you be able to deduce from the answer to the question whether the student has an insight into the concept being questioned?
- Do you have any comments on the formulation of the question or the clarity of the figures?

The experts pointed out the following issues:
- To stress the symmetry between the Sun and the star questions it is important that parallel questions are formulated using the same words;
- The jargon words in some questions should be avoided (e.g., culmination);

*TOAST test item 4*

You are located in the continental U.S. on the first day of October. How will the position of the Sun at noon be different two weeks later?
a) It will have moved toward the north.
b) It will have moved to a position higher in the sky.
c) It will stay in the same position.
d) It will have moved to a position closer to the horizon.
e) It will have moved toward the west.

*AMoSS question I.B*

On March 21st, an observer in Brussels sees the Sun at its highest point, as shown in the figure. Where does this observer see the Sun one month later at its highest point?
a) Near point a
b) Near point b
c) Near point c
d) Near point d
e) In the same point as on March 21st
f) I really don't know.

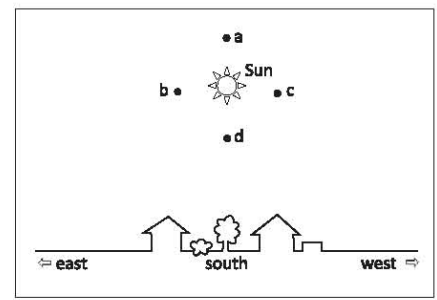

FIG. 2. Example of how test item 4 of the TOAST [13] is redesigned to AMoSS question I.B.

- The questions about the projected speed of the object on its path on the celestial sphere (category F), may be difficult for secondary school students;
- Some figures have to be redrawn.

Based on this feedback, we reformulated the questions.

### C. Pilot testing

Testing the AMoSS test was done in several steps. In a first pilot study, 42 students of a Flemish school took the test in their fifth year of secondary education (16–17 year olds). For these students, an introduction to astronomy is part of the geography curriculum. The rotation of the Earth about its axis, the Earth's revolution and the cause of seasons are discussed. All elements of the test were addressed in one way or another during the lessons, but the apparent celestial motions were not explicitly discussed.

The test was administered after this series of lectures during a science class. The students were free to decide whether or not to participate. No incentive was given to the students, but the science teacher motivated them to do an extra effort and he told them he would check the answers to be sure the students had taken the test seriously. Only two students had not finished when the science class, which lasted 45 min, ended. Only those two students returned an incomplete test: one student skipped one question, but the other student skipped 7 questions.

The collected answers were analyzed, and a score of 1 was given if the correct alternative was chosen and 0 if an incorrect alternative was chosen or if no answer was given. We present the results in Table II. The mean score for all participants on all questions was $M = 40\%$, $SD = 14\%$ ($n = 42$). On average the seven Sun questions ($M = 46\%$, $SD = 20\%$) are answered more correctly than the seven star questions ($M = 34\%$, $SD = 15\%$). Figure 3 shows a more detailed view and reveals that very different scores were obtained over the different questions. Figure 3 is organized by category (A, B, …), to reflect the symmetry between the Sun and the star questions, as presented in Table I. On questions about the influence of the observer's position on the Sun star-rise set (category D), the speed of the apparent motion of Sun stars (category E), and about seasons and sky map changes (category III and IV), students scored better on the star questions than on the Sun questions. For all other questions, the Sun questions are better answered than the stellar analogue.

TABLE II. Test results of secondary school students ($n = 42$).

|  | Questions | | |
| --- | --- | --- | --- |
|  | Sun | Star | Total |
| Average | 46% | 34% | 40% |
| Median | 43% | 29% | 43% |
| Standard deviation | 20% | 15% | 14% |

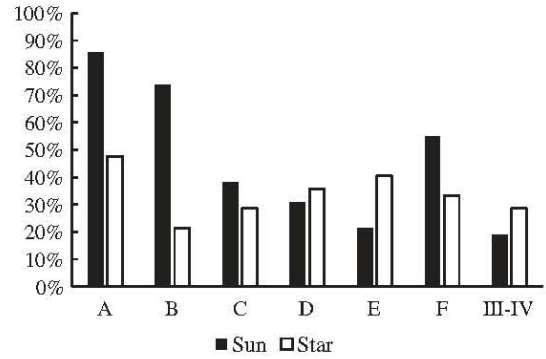

FIG. 3. Percentage of the secondary school students with a correct answer on the Sun and star questions.

The written explanations were used to check whether the formulation of the questions was clear enough and whether the figures were interpreted correctly.

From the analysis of the scores and the written explanations, we concluded the following:

- The formulation of four questions had to be refined because students sometimes misinterpreted the question. For example, for "The Sun is high in the south" some students thought that the observer was in the southern hemisphere of the Earth. We therefore mentioned in the question in which city the observer was located.
- There seemed to be a problem with the category F questions (see Fig. 4): none of the written explanations was correct, even when the correct answer was ticked.
- There was a lot of guessing. Some students wrote this literally in the explanation of their answer. The distribution of results on some questions also supports this suspicion. To reduce noise on the data, in the next version of the test, we added an alternative: "I really don't know."

The arc between point a and point b describes Denebola's motion in the sky for an observer in Brussels on March 21st between 22h00 and 24h00.

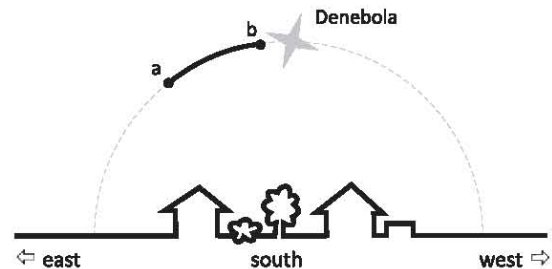

If you would redraw this figure for the same observer in Brussels for the star's motion on June 21st between 22h00 and 24h00, what can you say about the length of the arc between point a and point b?
a) The arc will be shorter than on March 21st.
b) The arc will have the same length as on March 21st.
c) The arc will be longer than on March 21st.

FIG. 4. Question II.F: None of the written explanations was correct, even when the correct answer was ticked.

In the next step we computed the item difficulty and item discrimination indices. The item difficulty for item $i$, $p_i$, is defined as the proportion of the participants who get that item correct [53]. Measuring item difficulty is useful in evaluating whether the difficulty of an item is suited to the level of participants administering the test. The item discrimination index indicates the degree to which responses to one item are related to responses to the other items in the test [53]. We determined item discrimination as a point-biserial correlation, which is calculated as a Pearson correlation between responses to a particular item and the total test scores. Item difficulty and item discrimination for each question are given in Table III. The values of the item difficulty vary between 0.19 and 0.86 with an average of 0.40. The item discrimination values for almost all questions are between 0.16 and 0.52, which are acceptable values according to Allen and Yen [53]. A point biserial of 0.15 or higher is considered satisfactory. Only question II.F (see Fig. 4) had a value smaller than 0.15: the negative value indicates that students who performed well on the test as a whole tended to miss this question and students who did not perform as well on the test as a whole got it right. This confirms the conclusion from the analysis of the written explanations that the category F questions should be removed: The concept of apparent speed on the trajectory and how this could change, is too difficult for the participating students.

In the second pilot study, 48 university students took the revised version of the test. The test was administered during the last lesson of the course "Introduction to Astronomy" of a Flemish university, but the topic of the test was not explicitly addressed in that course. The students, too, could voluntarily decide whether or not to participate and no incentive was given. We asked again to justify the choice made for each question. Now the students also had the opportunity to indicate "I really don't know." We explicitly asked the students not to guess. The test lasted 60 min. All 48 students submitted the test, 9 students indicated not to have enough time to answer all the questions.

We analyzed the answers and calculated the scores in the same way as in the first pilot study. The results are presented in Table IV. The mean score for all participants on all questions was $M = 50\%$, $SD = 16\%$ ($n = 48$), which is a higher score than the secondary school students. On average the seven Sun questions ($M = 57\%$, $SD = 16\%$) are answered more correctly than the seven star questions ($M = 44\%$, $SD = 24\%$). Figure 5 with a comparison of the Sun and star questions per category, gives a more detailed view and reveals that most Sun questions are better answered than the star questions.

From the analysis of the scores and the written statements, we concluded the following:
- The formulation of questions I.D and II.D still needs to be adapted because it was not mentioned in which hemisphere the unknown cities were located. This was solved by using cities that are well known by the participants.
- We identified the same problem with questions of category F as in the first pilot study. Therefore, we omitted these two questions in the next version of the test and also omitted the corresponding category from the proposed framework (Table I). This leaves 6 categories and 12 questions.

We again calculated item difficulty and item discrimination indices for this test group (Table V). The values of the item difficulty vary between 0.19 and 0.96 with an average of 0.50, which is described as ideal by Allen and Yen [53]. The values of the item discrimination vary for most questions between 0.20 and 0.61. Only question I.C has a lower value of 0.10.

TABLE III. Item difficulty and item discrimination of the first pilot study (secondary school students $n = 42$).

|  | Item difficulty | Point biserial |
|---|---|---|
| Question I.A | 0.86 | 0.34 |
| Question II.A | 0.48 | 0.18 |
| Question I.B | 0.74 | 0.16 |
| Question II.B | 0.21 | 0.20 |
| Question I.C | 0.38 | 0.52 |
| Question II.C | 0.29 | 0.30 |
| Question I.D | 0.31 | 0.52 |
| Question II.D | 0.36 | 0.21 |
| Question I.E | 0.21 | 0.35 |
| Question II.E | 0.40 | 0.40 |
| Question I.F | 0.55 | 0.41 |
| Question II.F | 0.33 | −0.06 |
| Question III | 0.19 | 0.39 |
| Question IV | 0.29 | 0.35 |

TABLE IV. Test results of university students ($n = 48$).

|  | Questions | | |
|---|---|---|---|
|  | Sun | Star | Total |
| Average | 57% | 44% | 50% |
| Median | 57% | 43% | 50% |
| Standard deviation | 16% | 24% | 16% |

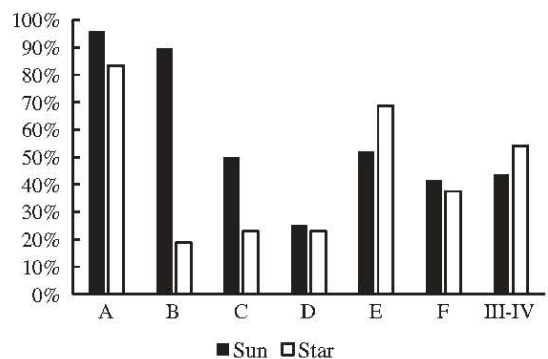

FIG. 5. Percentage of the university students with a correct answer on the Sun and star questions.

TABLE V. Item difficulty and item discrimination of the second pilot study (university students $n = 48$).

|  | Item difficulty | Point biserial |
|---|---|---|
| Question I.A | 0.96 | 0.20 |
| Question II.A | 0.83 | 0.42 |
| Question I.B | 0.90 | 0.20 |
| Question II.B | 0.19 | 0.47 |
| Question I.C | 0.50 | 0.10 |
| Question II.C | 0.23 | 0.53 |
| Question I.D | 0.25 | 0.20 |
| Question II.D | 0.23 | 0.26 |
| Question I.E | 0.52 | 0.29 |
| Question II.E | 0.69 | 0.39 |
| Question I.F | 0.42 | 0.30 |
| Question II.F | 0.38 | 0.61 |
| Question III | 0.44 | 0.49 |
| Question IV | 0.54 | 0.54 |

TABLE VI. Test results of secondary school students ($n = 33$).

| | Questions | | |
|---|---|---|---|
| | Sun | Star | Total |
| TEST | | | |
| Average | 53% | 34% | 43% |
| Median | 50% | 33% | 50% |
| Standard deviation | 26% | 21% | 20% |
| RETEST | | | |
| Average | 66% | 33% | 49% |
| Median | 67% | 33% | 50% |
| Standard deviation | 24% | 23% | 19% |

In the next step of the validation process, the questionnaire was again presented to a panel of experts, consisting of 8 professors and the director of the Brussels planetarium, each with their own expertise in drawing up a questionnaire for scientific research: 4 professors in science education, 2 professors in educational psychology, and 2 professors in astronomy. The expert panel asked for minor adjustments to the figures of some questions.

### D. Test-retest reliability

To finalize the validation process, we have checked the test-retest reliability by administering the final version of the AMoSS test with a group of 33 students of the 5th year of a Flemish secondary school (16–17 year olds) at two different points in time. The first point was chosen immediately after finishing the cosmography lessons, just before the Christmas exams. The second was chosen 7 weeks later. In between there were no science lessons concerning the topic of the apparent motion of the Sun and stars. During these weeks, students had to do exams and took Christmas holidays.

The results of these two tests are presented in Table VI. The mean score for all participants on all questions of the first test was $M = 43\%$, $SD = 20\%$ ($n = 33$). For the retest the mean score was $M = 49\%$, $SD = 20\%$. The small improvement of the total score is due to an overall better score on the questions about the Sun. The mean score on the seven Sun questions improved from 52% to 66%. The exam preparation may have helped the students to score better on the Sun questions.

With a Pearson correlation coefficient of 0.70, we can regard the test-retest reliability as acceptable.

## IV. DISCUSSION AND CONCLUSION

We developed and validated a test with 12 multiple choice questions (see the Supplemental Material [52] for the complete questionnaire) of which each question focuses on a specific element of the apparent motion of the Sun and stars (AMoSS) (see Table I). The development of this test is the first step in our study of opportunities of planetarium visualizations for the learning of these apparent motions. During the design of this test we paid explicit attention to the parallel formulation between questions relating to the apparent motion of the Sun and stars: for each element there is one question about the Sun and an almost identical question about the stars (see Table I).

The item difficulty and discrimination indices of the 12 test items were calculated. The item difficulty indices span a wide range which exceeds the range between 0.30 and 0.70, which Allen and Yen [53] describe as ideal to maximize the information a test provides about differences among respondents. However, we have decided to keep questions with a high difficulty index (e.g., 0.96) or a rather low difficulty index (e.g., 0.19). One of the aims of the AMoSS test is to grasp how different students answer the Sun and the parallel star questions. Therefore, to keep the symmetry, also too easy or too difficult questions have an important function in the questionnaire.

For the two pilot groups the item difficulty has an average of 0.40 for the secondary school students and an average of 0.50 for the university students. These overall results confirm that students often have an unclear understanding of the apparent motion of celestial bodies, as described in literature [13,48,49].

Similar to earlier studies [14,15,17,18], the results of the pilot test reveal that the test items on the apparent motion of the Sun are in general answered better than items on the apparent motion of stars. While prior studies [3,9,37,55–57] about apparent celestial motions are mostly limited to the Sun or the Moon and in few cases to certain elements of the Sun's path and star trail (e.g., culmination height), the AMoSS test allows to examine in more detail and more systematically which elements of the apparent motion of Sun and stars students really understand. By the symmetry between Sun questions and star questions, we are able to compare students' understanding of both apparent motions. Because each element is questioned separately and we ask for student reasoning, we can get very specific information about the knowledge and insights of the students and look for patterns in their answers across different categories. In this manuscript we have not reported on the examination of the mental models students use while explaining their answers. We reserve this for future work.

Although the validation shows we designed a useful test, we are aware of two possible limitations. First, to address a possible weakness of multiple choice testing and avoid student guessing, we have added an alternative "I really don't know" to most of the questions. When using the AMoSS test in a pretest and post-test setting the analysis of how students change their answer to or from "I don't know" will yield valuable information about the educational activity in between the pre- and the post-test [58]. Still, the multiple choice format enables students to choose or guess an answer rather than formulating an answer based on a physical reasoning.

Second our findings may be dependent on the Flemish educational context. In Flanders, students only start a systematic study of basic astronomical concepts in the fifth year of secondary school (16–17 year olds). The AMoSS test is developed for this target group and in the questions typical Flemish cities are used as locations for the observer of Sun and stars. However, the questions can easily be adapted to other countries, even to other age groups.

By applying our test on a larger scale, we will not only probe the overall understanding of these apparent motions, but also map what range of alternative ideas prevail. By examining the student's reasoning on a larger scale, we aim to identify the common alternative ideas in students' reasoning about the apparent motion of Sun and stars. Since all items of the AMoSS test are multiple choice questions, the analysis of how the different distractors are selected by the students can reveal which alternative ideas students have about the apparent motion of the Sun and stars.

In future work we will use the AMoSS test to measure to what extent a planetarium visit in the context of a school trip supports students' insight in the apparent motion of the Sun and stars. In collaboration with the Brussels planetarium we will take the test just before and shortly after their presentation and map out on which points students' insight has changed by following the presentation. This allows us to measure the effectiveness of this planetarium presentation on a cognitive level. Planetarium education research indicates that on this level the classical planetarium presentation is not necessarily more successful than a classical astronomy lesson at school [19]. Our ultimate aim is to design a research based planetarium session to maximize the learning effect.

We see also opportunities to use this—possibly adapted version of the—test in different age groups: educators can gauge the insight into the apparent motion of the Sun and stars among primary school children, secondary school students, and university students. It could be very interesting to compare these age groups in order to assess the impact of education on the results.

## ACKNOWLEDGMENTS

The preparation of this test is part of the Ph.D. research carried out by H. B. in collaboration with the other authors, all of whom are involved in the Studying Learning Opportunities in a Planetarium Environment (SLOPE) project. The authors would like to thank the Brussels planetarium staff for their advice and all pupils, students, teachers and professors who participated in this study. Also, thanks to the anonymous reviewers for their constructive suggestions in how to improve this manuscript.

[1] J. D. Plummer and J. Krajcik, Building a learning progression for celestial motion: Elementary levels from an earth-based perspective, J. Res. Sci. Teach. **47,** 768 (2010).

[2] J. D. Plummer, K. D. Wasko, and C. Slagle, Children learning to explain daily celestial motion: Understanding astronomy across moving frames of reference, Int. J. Sci. Educ. **33,** 1963 (2011).

[3] I. Testa, S. Galano, S. Leccia, and E. Puddu, Development and validation of a learning progression for change of seasons, solar and lunar eclipses, and moon phases, Phys. Rev. ST Phys. Educ. Res. **11,** 020102 (2015).

[4] R. Trumper, University students' conceptions of basic astronomy concepts, Phys. Educ. **35,** 9 (2000).

[5] J. M. Bailey and T. F. Slater, A review of astronomy education research, Astron. Educ. Rev. **2,** 20 (2003).

[6] A. Lelliott and M. Rollnick, Big ideas: A review of astronomy education research 1974-2008, Int. J. Sci. Educ. 2010) 1771 ,**32**).

[7] J. Wilhelm, M. Cole, C. Cohen, and R. Lindell, How middle level science teachers visualize and translate motion, scale, and geometric space of the Earth-Moon-Sun system with their students, Phys. Rev. Phys. Educ. Res. **14,** 010150 (2018).

[8] A. T. M. Mcdevitt, J. E. Ormrod, G. Cupit, M. Chandler, and V. Aloa, *Child Development and Education* (Pearson Higher Education AU, Pearson Australia, 2012).

[9] S. Vosniadou and I. Skopeliti, Is it the Earth that turns or the Sun that goes behind the mountains? students' misconceptions about the day/night cycle after reading a science text, Int. J. Sci. Educ. **39**, 2027 (2017).

[10] S. Vosniadou, The development of students' understanding of science, Front. Educ. **4**, 1 (2019).

[11] S. Carey, *Conceptual Change in Childhood* (The MIT Press, Cambridge, MA, 1986).

[12] D. F. Treagust, M. Won, and F. McLure, *Converging Perspectives on Conceptual Change* (Routledge, London, 2017).

[13] S. J. Slater, Sharon Price Shleigh, and D. J. Stork, Analysis of individual Test of Astronomy STandards (TOAST) item responses, J. Astron. Earth Sci. Educ. **2**, 89 (2015).

[14] J. D. Plummer, A cross-age study of children's knowledge of apparent celestial motion, Int. J. Sci. Educ. **31**, 1571 (2009).

[15] S. Vosniadou and W. F. Brewer, Mental models of the day/night cycle, Cogn. Sci. **18**, 123 (1994).

[16] R. Trumper, A cross-age study of junior high school students' conceptions of basic astronomy concepts, Int. J. Sci. Educ. **23**, 1111 (2001).

[17] J. Mant and M. Summers, Some primary-school teachers' understanding of the Earth's place in the universe, Res. Pap. Educ. **8**, 101 (1993).

[18] J. Plummer, Students' development of astronomy concepts across time, Astron. Educ. Rev. **7**, 139 (2009).

[19] T. F. Slater and M. Ratcliffe, *Teaching Astronomy in the Planetarium* (Springer, New York, 2017).

[20] S. Vosniadou and W. F. Brewer, Mental models of the earth: A study of conceptual change in childhood, Cogn. Psychol. **24**, 535 (1992).

[21] C. Sneider, V. Bar, and C. Kavanagh, Learning about seasons: A guide for teachers and curriculum developers, Astron. Educ. Rev. **10**, 010103 (2011).

[22] B. M. C. Lopresto and S. R. Murrell, An astronomical misconceptions survey, J. Coll. Sci. Teach. **40**, 14 (2011), https://eric.ed.gov/?id=EJ963545.

[23] R. Trumper, Teaching future teachers basic astronomy concepts—Seasonal changes—At a time of reform in science education, J. Res. Sci. Teach. **43**, 879 (2006).

[24] R. Trumper, A cross-age study of junior high school students' conceptions of basic astronomy concepts, Int. J. Sci. Educ. **23**, 1111 (2001).

[25] C. Türk and H. Kalkan, Teaching seasons with hands-on models: Model transformation, Res. Sci. Technol. Educ. **36**, 324 (2018).

[26] R. K. Atwood and V. A. Atwood, Preservice elementary teachers' conceptions of the causes of seasons, J. Res. Sci. Teach. **33**, 553 (1996).

[27] D. Heywood, J. Parker, and M. Rowlands, Exploring the visuospatial challenge of learning about day and night and the sun's path, Sci. Educ. **97**, 772 (2013).

[28] K. C. Yu, K. Sahami, V. Sahami, and L. C. Sessions, Using a digital planetarium for teaching seasons to undergraduates, J. Astron. Earth Sci. Educ. **2**, 33 (2015).

[29] M. Cole, C. Cohen, J. Wilhelm, and R. Lindell, Spatial thinking in astronomy education research, Phys. Rev. Phys. Educ. Res. **14**, 010139 (2018).

[30] L. Zwartjes, M. L. de Lázaro, K. Donert, I. Buzo, R. de Miguel Gonzalez, and E. Woloszynska-Wisniewska, Literature review on spatial thinking, GI Learner Project (2017), http://hdl.handle.net/1854/LU-8623271.

[31] J. D. Plummer, Spatial thinking as the dimension of progress in an astronomy learning progression, Stud. Sci. Educ. **50**, 1 (2014).

[32] U. Eriksson, Disciplinary discernment: Reading the sky in astronomy education, Phys. Rev. Phys. Educ. Res. **15**, 010133 (2019).

[33] C. A. Cohen and M. Hegarty, Individual differences in use of external visualisations to perform an internal visual task, Appl. Cogn. Psychol. **21**, 701 (2007).

[34] Y. Kali and N. Orion, Spatial abilities of high-school students in the perception of geologic structures, J. Res. Sci. Teach. **33**, 369 (1996).

[35] M. Kozhevnikov and R. Thornton, Real-time data display, spatial visualization ability, and learning force and motion concepts, J. Sci. Educ. Technol. **15**, 111 (2006).

[36] J. D. Plummer, A. Kocareli, and C. Slagle, Learning to explain astronomy across moving frames of reference: Exploring the role of classroom and planetarium-based instructional contexts, Int. J. Sci. Educ. **36**, 1083 (2014).

[37] J. D. Plummer and L. Maynard, Building a learning progression for celestial motion: An exploration of students' reasoning about the seasons, J. Res. Sci. Teach. **51**, 902 (2014).

[38] K. C. Yu, K. Sahami, G. Denn, V. Sahami, and L. C. Sessions, Immersive planetarium visualizations for teaching solar system moon concepts to undergraduates, J. Astron. Earth Sci. Educ. **3**, 93 (2016).

[39] E. Lantz, Planetarium of the future, Curator Museum J. **54**, 293 (2011).

[40] R. S. Lindell, E. Peak, and T. M. Foster, Are they all created equal? A comparison of different concept inventory development methodologies, AIP Conf. Proc. **883**, 14 (2007).

[41] R. S. Lindell and J. P. Olsen, Developing the lunar phases concept inventory, in *Proceedings of the 2002 Physics Education Research Conference, Boise, ID* (AIP, New York, 2002).

[42] E. M. Bardar, E. E. Prather, K. Brecher, and T. F. Slater, Development and validation of the light and spectroscopy concept inventory, Astron. Educ. Rev. **5**, 103 (2009).

[43] P. M. Sadler, H. Coyle, J. L. Miller, N. Cook-Smith, M. Dussault, and R. R. Gould, The Astronomy and Space Science Concept Inventory: Development and validation of assessment instruments aligned with the K–12 national science standards, Astron. Educ. Rev. **8**, 1 (2010).

[44] J. M. Bailey, B. Johnson, E. E. Prather, and T. F. Slater, Development and validation of the star properties concept inventory, Int. J. Sci. Educ. **34**, 2257 (2012).

[45] B. Hufnagel, Development of the astronomy diagnostic test, Astron. Educ. Rev. **1**, 47 (2009).

[46] T. F. Slater and S. J. Slater, Development of the Test of Astronomy Standards (TOAST) assessment instrument, Bull. Am. Astron. Soc. **40**, 273 (2008).

[47] V. Rajpaul, S. Allie, and S. L. Blyth, Introductory astronomy course at the University of Cape Town: Probing student perspectives, Phys. Rev. ST Phys. Educ. Res. **10**, 020126 (2014).

[48] M. C. LoPresto, Astronomy diagnostic test results reflect course goals and show room for improvement, Astron. Educ. Rev. **5**, 16 (2009).

[49] B. Hufnagel, T. Slater, G. Deming, J. Adams, R. L. Adrian, C. Brick, and M. Zeilik, Pre-course results from the astronomy diagnostic test, Pub. Astron. Soc. Aust. **17**, 152 (2000).

[50] W. K. Adams and C. E. Wieman, Development and validation of instruments to measure learning of expert-like thinking, Int. J. Sci. Educ. **33**, 1289 (2011).

[51] American Educational Research Association, American Psychological Association, National Council on Measurement in Education, and Joint Committee on Standards for Educational and Psychological Testing (U.S.), *Standards for Educational, and Psychological Testing* (American Educational Research Association, New York, 2014).

[52] See Supplemental Material at http://link.aps.org/supplemental/10.1103/PhysRevPhysEducRes.16.020135 for APPENDIX A: AMoSS QUESTIONNAIRE (Final version) and APPENDIX B: AMoSS questions.

[53] W. M. Allen and M. J. Yen, *Introduction to Measurement Theory* (Waveland Press. Inc., Long Grove, IL, 1997).

[54] W. R. Thornburgh, The role of the planetarium in students' attitudes, learning, and thinking about astronomical concepts, Electron. Theses Diss. Pap. 2684 (2017).

[55] S. Galano, A. Colantonio, S. Leccia, I. Marzoli, E. Puddu, and I. Testa, Developing the use of visual representations to explain basic astronomy phenomena, Phys. Rev. Phys. Educ. Res. **14**, 010145 (2018).

[56] R. Trumper, The need for change in elementary school teacher training—A cross-college age study of future teachers' conceptions of basic astronomy concepts, Teach. Teach. Educ. **19**, 309 (2003).

[57] C. M. Betts and A. Pattee, Flipping about the sun and its pattern of apparent motion, J. STEM Arts Crafts Constr. **2**, 8 (2017), https://scholarworks.uni.edu/journal-stem-arts/vol2/iss1/2/.

[58] K. Spears and M. Wilson, 'I don't know' and multiple choice analysis of pre- and post-tests, J. Ext. **48** (2010), https://www.researchgate.net/publication/266287703.